\documentclass[%
 aip,
 amsmath,amssymb,
 reprint,%
]{revtex4-2}

\usepackage{graphicx}
\usepackage{dcolumn}
\usepackage{xcolor}
\usepackage{float}

\usepackage[utf8]{inputenc}
\usepackage[T1]{fontenc}
\usepackage{mathptmx}
\usepackage{etoolbox}

\makeatletter
\def\@email#1#2{%
 \endgroup
 \patchcmd{\titleblock@produce}
  {\frontmatter@RRAPformat}
  {\frontmatter@RRAPformat{\produce@RRAP{*#1\href{mailto:#2}{#2}}}\frontmatter@RRAPformat}
  {}{}
}%
\makeatother

\usepackage{xspace}
\usepackage{amsmath}

\usepackage{multirow}
\usepackage{array}

\usepackage{lipsum}
\usepackage{subfigure}
\usepackage[justification=raggedright]{caption}

\makeatletter
\def\maketitle{
\@author@finish
\title@column\titleblock@produce
\suppressfloats[t]}
\makeatother

\begin{document}
\author{Rajarshi Sinha-Roy}%
\email{rajarshi.sinha-roy@univ-lyon1.fr}
\affiliation{ 
Université Lyon 1, CNRS, Institut Lumière Matière, UMR5306, F-69100, Villeurbanne, France
}%
\author{Cl\'ement Guiot du Doignon}
\affiliation{ 
Université Lyon 1, CNRS, Institut Lumière Matière, UMR5306, F-69100, Villeurbanne, France
}%
\author{Evan Munaro-Langloÿs}
\affiliation{ 
Université Lyon 1, CNRS, Institut Lumière Matière, UMR5306, F-69100, Villeurbanne, France
}%
\author{Franck Rabilloud}%
\affiliation{ 
Université Lyon 1, CNRS, Institut Lumière Matière, UMR5306, F-69100, Villeurbanne, France
}%
\author{Victor Despr\'e}%
\email{victor.despre@univ-lyon1.fr}
\affiliation{ 
Université Lyon 1, CNRS, Institut Lumière Matière, UMR5306, F-69100, Villeurbanne, France
}%

\title{
Real-time simulation of charge migration within the time-dependent Kohn-Sham DFT
}

\date{\today}

\begin{abstract}

Attosecond technologies provide unique opportunities to study electron dynamics and electron correlation on their intrinsic timescales. From a theoretical perspective, this places strong constraints as an accurate treatment of electron correlation is required. Recently, it was demonstrated that time-dependent density-functional theory (TDDFT) is capable of correctly predicting correlation-driven charge migration arising from hole mixing following ionization of the highest occupied molecular orbital (HOMO). Given the ability of TDDFT to treat large-scale systems, this approach offers promising perspectives for investigating electron-correlation-driven mechanisms in complex molecules. In this work, we assessed the constraints and limitations associated with using TDDFT to study this mechanism. We found that the charge-migration dynamics are already correctly reproduced using local-density approximation for the exchange-correlation functional, provided the states involved in the coherent superposition are well described within the TDDFT. However, for dynamics triggered by the ionization of orbitals below the HOMO, artificial ultrafast dynamics may appear on top of the charge-migration dynamics. These artifacts indicate that careful analysis of the simulated dynamics is required in order to reliably predict phenomena that could be observed experimentally.

\end{abstract}

\maketitle

The attosecond timescale corresponds to the characteristic time of electron motion at the atomic scale. The development of attosecond technologies has thereby enabled the study of electron dynamics with unprecedented time resolution \cite{antoine1996attosecond,krausz2009attosecond}. Achieving this time resolution allows for the precise investigation of electron correlation. Initially demonstrated in atoms and small molecules \cite{sansone2010electron,neidel2013probing}, over the past decade attosecond-timescale measurements have been extended to larger molecules \cite{calegari2014ultrafast,marciniak2019electron}, solids \cite{cavaletto2025attoscience,sederberg2020attosecond}, and liquids \cite{li2024attosecond}.

A major driving force behind the advancement of attosecond science has been the unveiling of fundamental mechanisms that were previously inaccessible \cite{lepine2014attosecond}. Since its inception, attosecond science has been guided and supported by theoretical predictions and simulations \cite{cederbaum1999ultrafast}. This presents a tremendous challenge for theorists, as it requires accurately modeling electron correlation and mechanisms that involve both high energy (often in the XUV range) and intense fields, owing to the intrinsic link between attosecond physics and high-intensity IR laser pulses.

Significant theoretical efforts have been made to study attosecond(as)-scale mechanisms. Various methods have proven successful, such as ADC \cite{kuleff2016core}, MCTDH \cite{matselyukh2022decoherence}, direct propagation \cite{kuleff2005multielectron}, R-Matrix \cite{feist2014time}, TD-SE \cite{sabbar2017state}, surface hopping \cite{grell2025modeling}, among others. As the field continues to focus on larger systems, time-dependent density functional theory (TDDFT)~\cite{RungeGross1984} emerges as an attractive method. TDDFT has demonstrated its ability to handle large systems, as evidenced by the extensive literature~\cite{Jornet-Somoza2015,Sinha-Roy2017ACSphotonics,Alvarez-Ibarra2020,Sinha-Roy2021JPCL,Mansson2022}. Its real-time formulation (RT-TDDFT)~\cite{YabanaBertsch1996} allows the study of nonlinear mechanisms, and in its real-space form, it can accurately simulate photoionization~\cite{Ulrich2012TDDFTbook}. Both processes are essential for attosecond science. However, as the TDDFT is employed through the Kohn-Sham scheme, where the effective time-dependent potential is a functional of the one-electron density, the excited states becomes single-determinant by nature. Therefore, Kohn-Sham TDDFT framework becomes unsuitable for simulating certain processes, particularly those involving multi-excitations \cite{kuleff2009theoretical}.

One of the mechanisms that has garnered considerable attention from both theoreticians~\cite{despre2015attosecond,mignolet2014charge,lara-astiaso2016decoherence,vacher2017electron,yuan2019ultrafast,folorunso2021molecular,zhang2024cavity,tremblay2026persistent,scheidegger2025can} and experimentalists~\cite{calegari2014ultrafast,lara-astiaso2018attosecond,kraus2015measurement,matselyukh2022decoherence,schwickert2022electronic,driver2024attosecond} in the attosecond community is charge migration. It refers to the ultrafast dynamics of the electron density of a molecule following ionization by a coherent, short pulse of light. Its correlation-driven variant~\cite{cederbaum1999ultrafast,kuleff2014ultrafast} refers to cases where features absent from mean-field electronic structure methods---such as configuration mixing or multiexcited configurations---are the primary drivers of charge migration. This mechanism holds fundamental importance as it arises solely from electron correlation, making it a direct probe of beyond-mean-field interactions between the constituent particles of matter. Notably, long-lived electron coherence have been predominantly predicted in the context of correlation-driven charge migration~\cite{despre2015attosecond,despre2018charge}, whereas ultrafast decoherence has been more commonly associated with other types of charge migration~\cite{vacher2017electron,arnold2017electronic}.

Correlation-driven charge migration can arise from three distinct electronic structure scenarios \cite{breidbach2003migration}: hole mixing, main states with satellites, and the breakdown of the molecular orbital picture \cite{cederbaum1986correlation}, which leads to the formation of a correlation band \cite{deleuze1996formation,herve2021ultrafast}. In the present work, we focus on the hole-mixing mechanism. Hole mixing occurs when cationic eigenstates cannot be adequately described by the removal of a single molecular orbital and instead require contributions from multiple ionization configurations. In this case, the cationic states are expressed as linear combinations of single-hole configurations corresponding to the ionization of at least two different orbitals. It has recently been shown that RT-TDDFT can successfully simulate this mechanism when the hole mixing involves a hole in the HOMO of a molecule \cite{du2025correlation}. In that study, it was demonstrated that infrared multiphoton ionization can trigger correlation-driven charge migration, opening the possibility of experimental investigation at free-electron laser facilities \cite{golubev2021core}.

To better assess the potential of RT-TDDFT for simulating correlation-driven charge migration arising from hole mixing, it is essential to evaluate how the choice of exchange-correlation functional affects the description of this mechanism. Although the DFT analog of Koopmans' theorem states that the negative of the Kohn–Sham HOMO energy approximates the molecular ionization potential \cite{casida1999correlated}, the accuracy with which Kohn–Sham orbitals describe ionization processes remains an important question. This is particularly relevant for ionization induced by short laser pulses, which can coherently populate ionic states separated by large energy gaps and thereby initiate ultrafast electron dynamics.

Correlation-driven charge migration in hole-mixing systems provides an ideal benchmark for addressing this question. Because these dynamics are generated solely by the ionization of a single orbital, they enable a direct assessment of how accurately Kohn–Sham orbitals represent the resulting electronic dynamics. Such studies will clarify the applicability of RT-TDDFT to ionization-induced charge migration, reduce the risk of predicting nonphysical dynamics, and ultimately increase confidence in using the method to guide and interpret experimental investigations.

\begin{figure}
\includegraphics[width=0.45\textwidth]{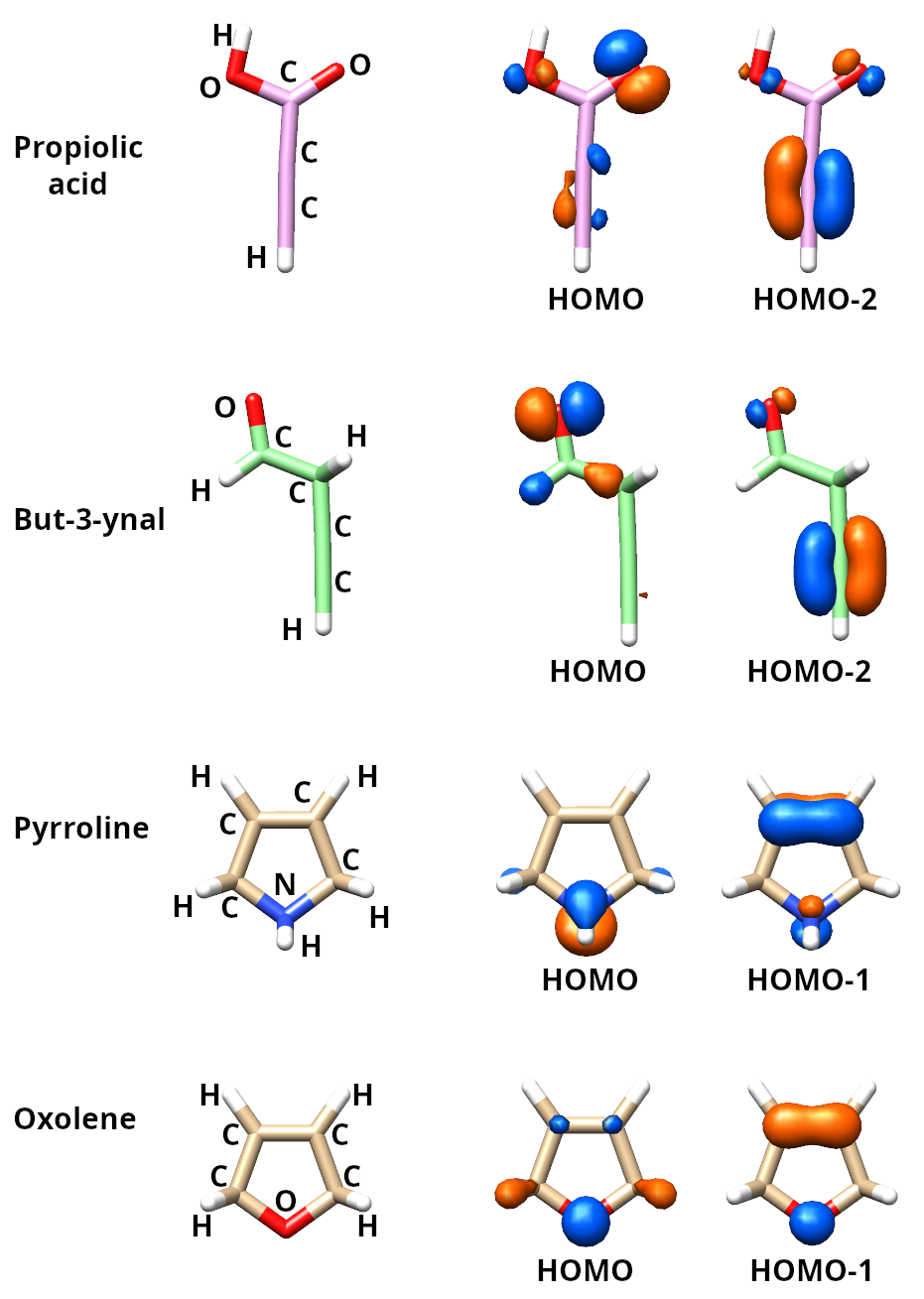}
\caption{
Geometry and the two low-lying Kohn-Sham orbitals that participate in the hole-mixing structures of the four molecules studied in this work. 
}
\label{molecules-HOMO-HOMO-n}
\end{figure}

In this work, we investigate correlation-driven charge migration in four molecules. These are propiolic acid, but-3-ynal, 3-pyrroline (hereinafter ``pyrroline''), and 2,5-dihydrofuran (hereinafter ``oxolene''). The geometries of the atomic arrangement in these four molecules are shown in Fig~\ref{molecules-HOMO-HOMO-n}. The four selected molecules exhibit hole mixing involving their HOMO \cite{despre2018charge,scheidegger2022search,du2025correlation}, meaning that ionization of the HOMO leads to a coherent superposition of two or more cationic eigenstates, resulting in ultrafast electronic beating that persists indefinitely if no dissipation effects are included. Previous studies have demonstrated that RT-TDDFT can successfully simulate correlation-driven charge migration arising from hole mixing involving their HOMO \cite{du2025correlation}. In all the four molecules the hole-mixing structure arises from strong contributions of the HOMO and another low-lying molecular orbital. These two orbitals calculated using local density approximation (LDA)~\cite{Kohn-Sham_LDA} are also shown for each of the four molecules in Fig~\ref{molecules-HOMO-HOMO-n}. While LDA is the simplest approximation in the Kohn-Sham DFT, in this work we test the gradient corrected PBE~\cite{Perdew1996_GGA_PBE} functional, and also consider an average-density based self-interaction correction (SIC)~\cite{Legrand2002} to show that even these relatively simple functionals are capable of correctly describing the dynamics of charge migration. At the same time, we show that ionization of low-lying orbitals can lead to non-physical behavior.

The simulations were performed using the open-source RT-TDDFT code Octopus \cite{andrade2015real,tancogne2020octopus}, which enables RT-TDDFT simulations in real space, i.e., on a numerical grid. The calculations were carried out on a spherical grid with a radius of 12~Å, a spacing of 0.18~Å, and a time step of 1.3~as. Absorbing boundary condition is employed in order to simulate the ionization process. The absorbing boundary is represented using a complex absorbing potential (CAP)~\cite{DeGiovannini2015} as described and implemented in the code Octopus. The CAP is defined by a sine-squared profile within an absorption layer padded at the inner wall of the computational domain. As a reference method, the third-order Algebraic Diagrammatic Construction approach, ADC(3) \cite{schirmer1998non,schirmer2018many}, was employed. ADC methods have been extensively used to describe correlation-driven charge migration and have provided reliable predictions \cite{kuleff2014ultrafast,despre2015attosecond,despre2018charge,lunnemann2008charge,scheidegger2025can}.

\begin{figure*}
\includegraphics[width=0.99\textwidth]{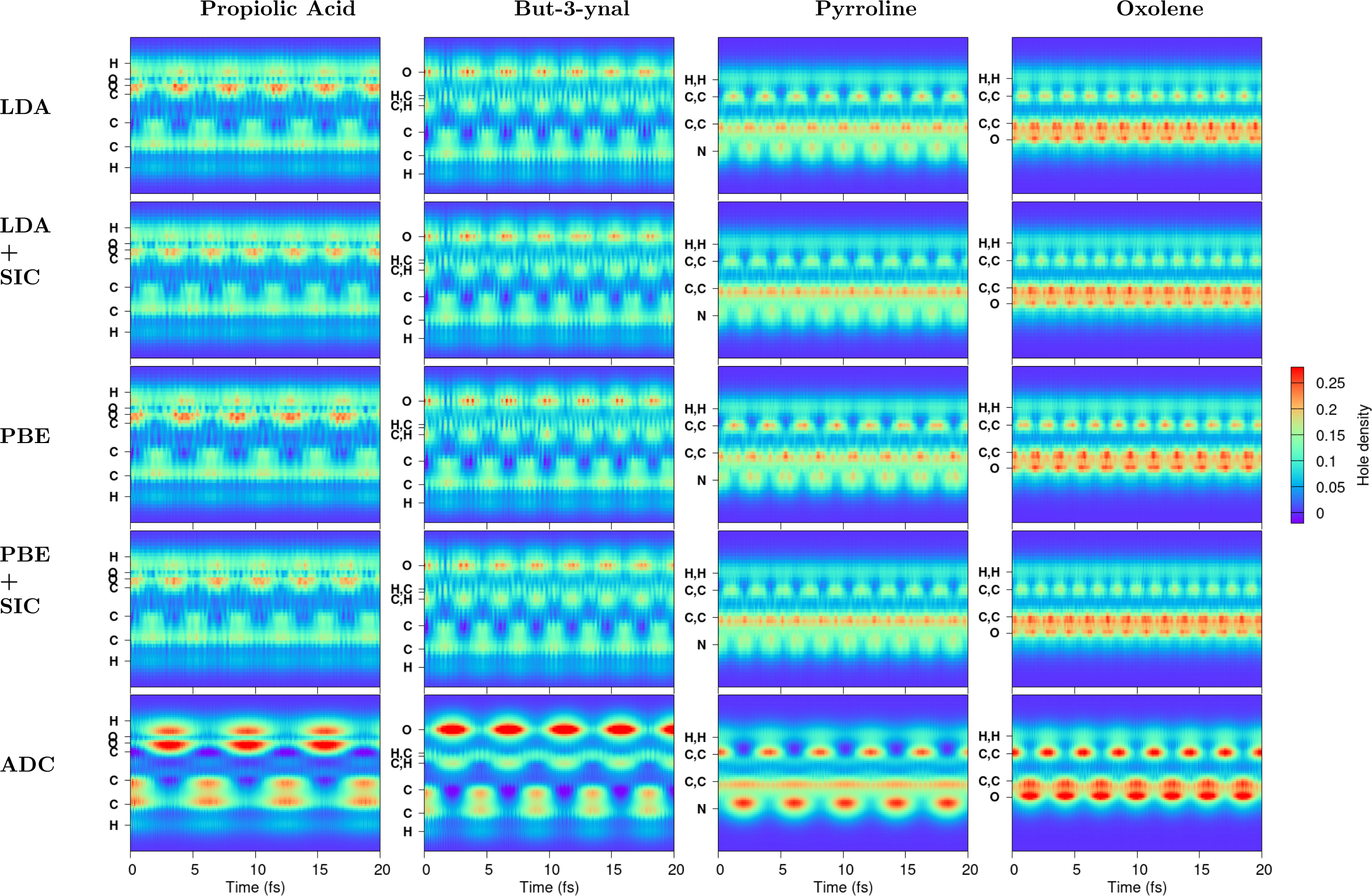}
\caption{
Color maps illustrating the time-dependent evolution of the hole density projected along the molecular axis after sudden ionization by the removal of an electron from the HOMO are presented for propiolic acid, but-3-ynal, pyrroline, and oxolene, from left to right, respectively.
For each of the molecules the results are shown for LDA, LDA+SIC, PBE, and PBE+SIC, from top to bottom, respectively. 
The values of the hole density are represented by the color bar. 
}
\label{SI-from-HOMO}
\end{figure*}

\begin{table*}

\centering
\captionof{table}{Frequency and period of charge-miration dynamics in different approximations.} \label{table1}

\setlength\extrarowheight{-2pt}

\begin{tabular}{ l c c c c c c c c c c}

\hline
Molecules & \multicolumn{2}{c}{LDA} & \multicolumn{2}{c}{LDA-SIC} & \multicolumn{2}{c}{PBE} & \multicolumn{2}{c}{PBE-SIC} & \multicolumn{2}{c}{ADC(3)}\\
& $\omega$ [eV] & $T$ [fs] & $\omega$ [eV] & $T$ [fs] & $\omega$ [eV] & $T$ [fs] & $\omega$ [eV] & $T$ [fs] & $\omega$ [eV] & $T$ [fs]\\
\hline
Propiolic acid 
& 1.099 & 3.76   
& 1.256 & 3.39 
& 0.942 & 4.39 
& 1.256 & 3.39 
& 0.67 & 6.2 \\
But-3-ynal 
& 1.413 & 2.92  
& 1.413 & 2.92
& 1.413 & 2.92 
& 1.413 & 2.92 
& 0.92 & 4.5 \\
3-Pyrroline 
& 1.649 & 2.507 
& 1.963 & 2.1 
& 1.649 & 2.507 
& 2.042 & 2.025 
& 1.03 & 4.0 \\
Oxolene 
& 2.356 & 1.755 
& 2.67 & 1.54 
& 2.199 & 1.87 
& 2.67 & 1.54 
& 1.65 & 2.5 \\
\hline 
\end{tabular}

\end{table*}
~\\

The dynamics of charge migration following ionization by a short laser pulse was calculated using the sudden ionization approximation. Within this approximation, an electron is removed from a molecular orbital at $t = 0$. Here, we aim to test the sensitivity of the resulting dynamics on the choice of exchange-correlation (XC) functional within the Kohn-Sham scheme by comparing the RT-TDDFT results with the ADC(3) reference. 
The results of the sudden ionization by removing an electron from the HOMO are presented in Fig.~\ref{SI-from-HOMO}.
Different levels of approximation for XC functional are compared against ADC(3) through the time-dependent hole density defined as $n_{_{GS}}(\mathbf{r})-n(\mathbf{r},t)$. Here $n_{_{GS}}(\mathbf{r})$ is the ground-state density calculated with static DFT and $n(\mathbf{r},t)$ is the time-dependent density followed by the sudden ionization at $t=0$. In Fig.~\ref{SI-from-HOMO}, the time-dependent hole density is shown along the molecular axis which is also the direction along which the ultrafast beating takes place in the respective molecules. In this representation, for a given time and for each point along the molecular axis, the hole-density is integrated over the cross-sectional plane perpendicular to the molecular axis. This integrated value is presented on the color map. The four molecules being almost planer, the projected position of the atoms along the molecular axis (i.e., the direction of the beating) is also presented as labels of the y-axes in all panels of the Fig.~\ref{SI-from-HOMO}.

At the bottom row of Fig.~\ref{SI-from-HOMO}, the ADC(3) simulations exhibit the expected quantum beating, which in these molecules corresponds to charge oscillations between the double or triple carbon bond and a nitrogen or oxygen atom. Both functionals, with or without the SIC, successfully reproduce the charge dynamics. For all four molecules, our previous study had already demonstrated that PBE+SIC provides an accurate description of the dynamics \cite{du2025correlation}. It is therefore noteworthy that even the simplest functional considered here, namely LDA, is capable of reproducing the mechanism in all four systems.

This observation raises the question of whether the description improves with increasing functional complexity. To address this point, we compared the oscillation periods obtained with RT-TDDFT to those predicted by ADC(3). The results are presented in Table~\ref{table1}. No clear trend emerges, indicating that the oscillation period, which is determined by the energy difference between the states involved in the coherent superposition, is not strongly affected by the choice of functional. In particular, the self-interaction correction does not significantly influence the dynamics and even leads to slightly poorer agreement with the ADC(3) results.

This behavior is not unexpected. Self-interaction corrections primarily affect the long-range behavior of the electronic potential. Moreover, within the average-density SIC scheme employed here, they can slightly deteriorate the accuracy of the potential at short range. This arises because, in the average-density SIC scheme, the correction is determined by the average electron density instead of the density associated with each individual orbital. Since the states participating in the coherent superposition are well localized around the molecular backbone, the self-interaction correction does not substantially improve their description and may therefore lead to somewhat poorer agreement with ADC(3). Nevertheless, it is important to emphasize that if the dynamics were triggered by an explicit laser pulse, as demonstrated in our previous work, the inclusion of a self-interaction correction would become essential to obtain accurate ionization potentials and to properly describe the outgoing photoelectron.

At this stage, we would like to discuss two important points. First, these results show that correlation-driven charge migration, although fundamentally governed by electron correlation, is already well described at the LDA level of theory, implying relatively limited constraints on TDDFT for capturing this mechanism. However, it should be emphasized that correlation-driven charge migration arises from the coherent superposition of cationic eigenstates. Consequently, if the superposition involves states that are known to be problematic for TDDFT, such as charge-transfer states, then more sophisticated functionals may be required to accurately describe the underlying physics. In the molecules studied here, the relevant states do not exhibit such pathological behavior, and therefore LDA proves to be sufficient.

Second, the initial localization of the hole differs between the ADC and RT-TDDFT simulations. This discrepancy arises because Hartree–Fock orbitals are used in ADC, whereas Kohn–Sham orbitals are employed in RT-TDDFT. As a result, the two levels of theory predict different sites for the first ionization. This is a direct consequence of hole mixing, which occurs when different ionization sites are nearly energetically equivalent, leading to delocalized cationic states and Dyson orbitals \cite{von1982hole}.

\begin{figure}
\includegraphics[width=0.495\textwidth]{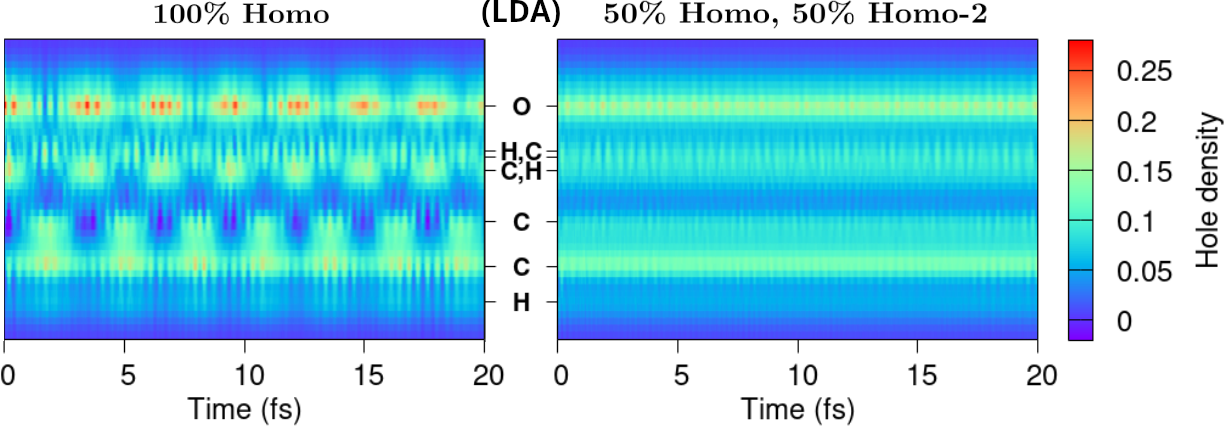}
\caption{
Sudden ionization by the removal of HOMO compared to the removal of 50\% of HOMO and 50\% HOMO-2 in Propiolic acid using LDA.
}
\label{50-50}
\end{figure}

This interpretation can be verified by performing a simulation in which the initial state is prepared as a coherent superposition in which the hole is created by removing half an electron from each of the orbitals involved in the hole mixing. In this case, the initial hole is delocalized over the molecule and closely resembles the corresponding Dyson orbital. As a consequence, a single cationic eigenstate is predominantly populated, resulting in the absence of charge migration dynamics. This behavior is observed in Fig.~\ref{50-50}, where such an initial hole in propiolic acid leads to a stationary delocalized positive charge distribution.

The difference in the initial hole localization is not critical for the prediction of experimental observables, since the actual localization of the hole depends on the characteristics of the ionizing pulse. Within the sudden ionization approximation, the most important requirement is the correct prediction of the quantum beating dynamics arising from the coherent superposition of cationic states.

\begin{figure*}
\includegraphics[width=0.99\textwidth]{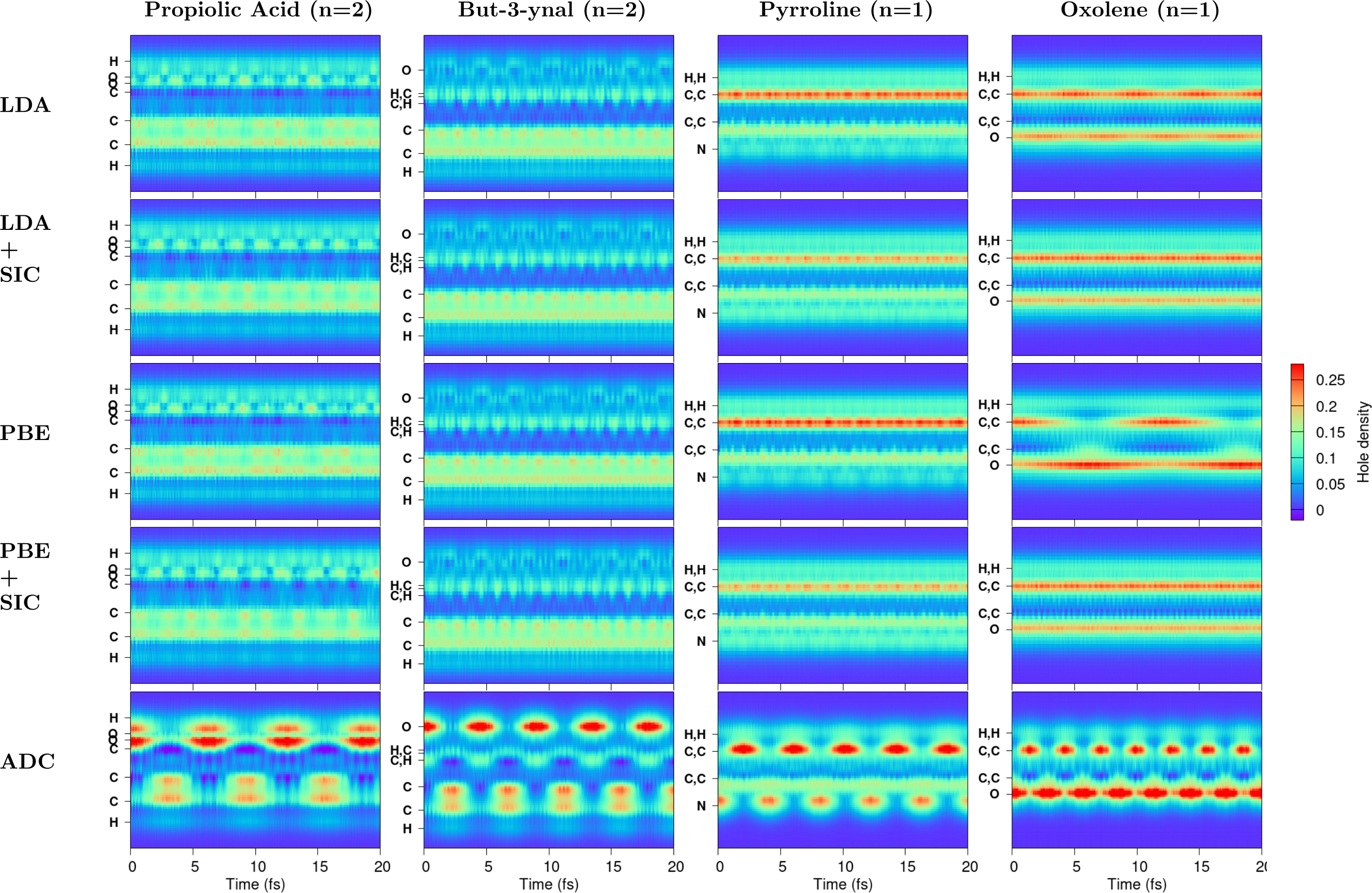}
\caption{
Color maps illustrating the time-dependent evolution of the hole density projected along the molecular axis after sudden ionization by the removal of an electron from the HOMO-n orbital are presented for propiolic acid (n=2), but-3-ynal (n=2), pyrroline (n=1), and oxolene (n=1), from left to right, respectively.
For each of the molecules the results are shown for LDA, LDA+SIC, PBE, and PBE+SIC, from top to bottom, respectively.
}
\label{SI-from-HOMO-n}
\end{figure*}

We would now like to assess how well RT-TDDFT describes hole mixing when a Kohn-Sham orbital other than the HOMO is ionized, i.e., outside the regime where Koopmans’ theorem hold. Following ADC(3) calculations, the orbitals that participate in the hole-mixing are the HOMO and the HOMO-2 for propiolic acid and but-3-ynal, and the HOMO and the HOMO-1 for pyrroline and oxolene. 
As a consequence, the dynamics due to ionization from the HOMO can also be produced by removing an electron from the other orbital that participates in hole mixing. We therefore simulated the ionization due to electron removal from the orbital other than the HOMO that participate in hole-mixing. The results obtained for the different XC approximations in TDDFT and using ADC(3) are presented in Fig.~\ref{SI-from-HOMO-n}.

\begin{figure}
\includegraphics[width=0.495\textwidth]{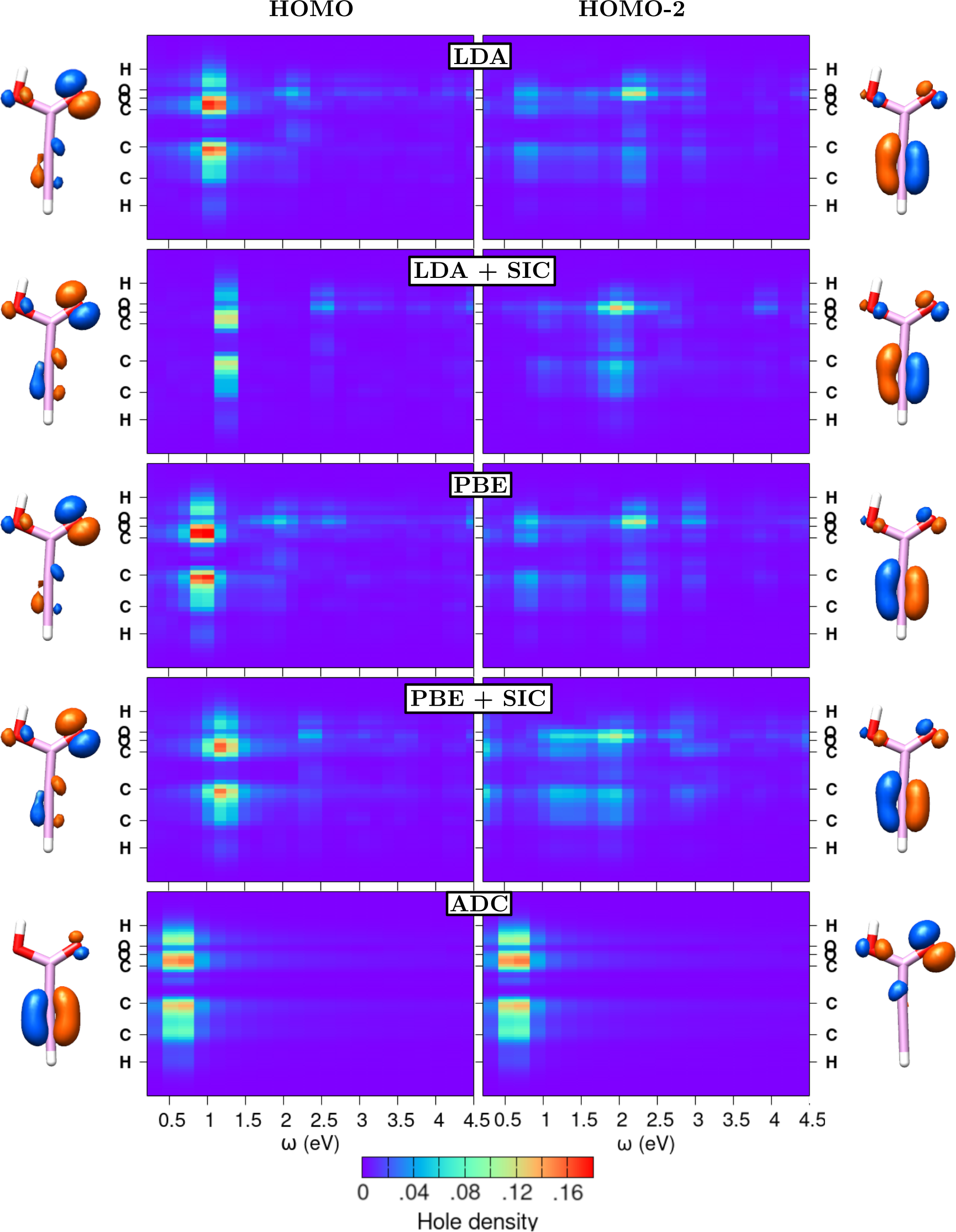}
\caption{
Fourier transform of the time-dependent density shows energy dependence of the charge dynamics following sudden ionization by removal of an electron from the HOMO (left) and HOMO$-2$ (right) of propiolic acid at different levels of theory. For RT-TDDFT, Kohn--Sham orbitals are used, whereas Hartree--Fock orbitals are employed in ADC(3). The corresponding molecular orbitals are shown for HOMO at the left and for HOMO$-2$ at the right.
}
\label{FT}
\end{figure}

As expected, ADC(3) correctly predicts dynamics similar to those obtained following HOMO ionization, with an inverted initial localization of the hole. In contrast, the RT-TDDFT simulations display noticeably different dynamics. In order to better understand the nature of the dynamics, in Fig.~\ref{FT} we plotted the frequency dependence by performing a discrete Fourier transformation of the dynamics obtained in Fig.~\ref{SI-from-HOMO} and Fig.~\ref{SI-from-HOMO-n} for the propiolic acid. The same frequency analyses for the other molecules are put in the supplementary material (SM). The HOMO and the HOMO-2 orbitals which participate in the hole-mixing structure of the propiolic acid are also presented for the different levels of theory at the left and the right side of the Fig.~\ref{FT} respectively. 
The resemblance of the HOMO in different approximations confirms the robustness of the charge-migration dynamics due to electron removal from HOMO. It is also noteworthy that the ordering of the HOMO and HOMO–2 in Kohn–Sham DFT is reversed relative to the Hartree–Fock (HF) orbitals used in the ADC(3) calculations. A similar interchange is observed in calculations based on the exact-exchange potential (see Fig.~4 in the SM). This confirms the different sites of initial localization of the hole in different levels of theory as mentioned earlier.

The frequency analysis shown in Fig.~\ref{FT} reveals that for the sudden ionization of the HOMO-2 of propiolic acid, the expected oscillation frequency, i.e., the frequency of oscillation observed for the sudden ionization of the HOMO, is present at all levels of approximation for the many-body description. However, an oscillatory hole dynamics of higher frequency is also observed. Both dynamics are also observed in but-3-ynal and pyrroline following the sudden ionization of orbitals other than the HOMO that participate in the hole mixing (cf. SM). The exception is oxolene, for which different dynamical behaviors are obtained depending on the exchange-correlation (XC) functional employed (cf. Fig.~3 in the SM). This suggests that the electronic structure of this system is more complex and, as discussed earlier, higher-level functionals may be required to accurately describe the relevant states. For other molecules, however, the expected dynamics is masked by additional ultrafast beatings that obscure the anticipated charge migration dynamics (cf. Fig.~\ref{FT}). Since Fig.~\ref{SI-from-HOMO} demonstrated that RT-TDDFT correctly describes the cationic states involved in the coherent superposition, these results indicate that the primary source of the discrepancy lies in the different molecular orbitals used to construct the ionized state.

To understand this behavior, it is necessary to consider the sudden ionization process in more detail. During sudden ionization, cationic eigenstates containing a single-electron component associated with the ionized orbital become populated, with the population amplitude determined by the weight of this component. Unlike ADC, which describes excited states as linear combinations of single, double, and higher-order excitations, TDDFT represents excited states within the single-determinant Kohn-Sham framework, where the currently-used local approximations for the XC functionals account only for single excitations. As a result, the contribution of single-electron configurations to highly excited states can be significantly overestimated within TDDFT. During sudden ionization, this can artificially generate coherent superpositions between states separated by large energy gaps, leading to ultrafast beatings. Although such superpositions are not impossible in principle, they are strongly overestimated in this case and may therefore result in nonphysical predictions. 

However, this also suggests a possible strategy for mitigating such nonphysical effects. By carefully designing the ionizing pulse, it may be possible to use a spectral bandwidth that is broad enough to coherently populate the cationic states involved in the hole-mixing mechanism, while remaining narrow enough to avoid populating the higher-lying states responsible for the artificial ultrafast beatings. Such pulse shaping could therefore help isolate the physically relevant charge migration dynamics and improve the reliability of RT-TDDFT predictions.

Therefore, when using TDDFT to model attosecond dynamics following ionization, particular care must be taken in assessing the predictive reliability of the simulations, especially when ultrafast beatings are observed.

In this work, following the demonstration that RT-TDDFT can be used to model correlation-driven charge migration arising from hole mixing after ionization of the HOMO, we sought to provide guidance on how sensitive the description of this mechanism is to the choice of simple local exchange-correlation functional and to evaluate the robustness of the predictions obtained with this approach.

Our results show that even the LDA functional is capable of describing hole mixing correctly. This indicates that the main constraint on the TDDFT functional does not arise from the hole-mixing mechanism itself, but rather from the accurate description of the electronic states involved in the superposition, which may require more advanced functionals depending on the system under investigation.

By triggering the dynamics through the ionization of low-lying orbitals, we observed that RT-TDDFT can produce artificial dynamics. These artifacts originate from two main factors: the nature of Kohn-Sham orbitals and the single-determinant character of excited states calculated within the Kohn-Sham framework using local XC functionals. The resulting nonphysical ultrafast beatings appear superimposed on the expected dynamics. Nevertheless, with careful analysis, RT-TDDFT may still be used to predict dynamics triggered by the ionization of orbitals beyond the HOMO.

These results are promising for the application of RT-TDDFT to attosecond science and open the possibility of extending such studies to systems whose size exceeds the capabilities of more computationally demanding methods such as ADC. These developments are particularly timely, as experimental investigations are increasingly focusing on larger and more complex systems \cite{zhou2025state,moore2025solvation,zhang2025intermolecular}, thereby creating a growing need for theoretical approaches capable of treating such scenarios \cite{munaro2026tailoring}. We hope that this work will stimulate further research in this direction and contribute to the establishment of practical guidelines for assessing the robustness and the reliability of predicted ultrafast electron dynamics.

\section*{SUPPLEMENTARY MATERIAL}

\section*{Acknowledgments.} 
R.S.-R. acknowledges support from the Université Lyon~1  in the frame of the programme ``Accueil Enseignants-Chercheurs'' AEC-2023, AEC-2024, and AEC-2025. Support from the GENCI-IDRIS (Grant 2023-0906829, Grant A0170807662, Grant A0190807662) and the PSMN (Pôle Scientifique de Modélisation Numérique) for high performance computing is gratefully acknowledged. Moreover, the authors acknowledge the contribution of the International Research Network IRN Atto-DICES (CNRS). 

\section*{AUTHOR DECLARATIONS}
\subsection*{Conflicts of interest}
There are no conflicts to declare.
\subsection*{Author contributions}
V. D. conceptualised the work with R.S.-R. Calculations, analysis of results, and preparing the figures are performed by C. G. D., E. L., and R. S. -R. The initial draft of the manuscript was prepared by R. S. -R. and V. D. All authors contributed to reviewing and editing the final manuscript and approved the final submitted version. 

\section*{Data availability}
The data that support the findings of this study are available within the article and its supplementary material.

\bibliography{RT-TDDFT.bib}

\end{document}


\author{Rajarshi Sinha-Roy}%
\email{rajarshi.sinha-roy@univ-lyon1.fr}
\affiliation{ 
Université Lyon 1, CNRS, Institut Lumière Matière, UMR5306, F-69100, Villeurbanne, France
}%
\author{Cl\'ement Guiot du Doignon}
\affiliation{ 
Université Lyon 1, CNRS, Institut Lumière Matière, UMR5306, F-69100, Villeurbanne, France
}%
\author{Evan Munaro-Langloÿs}
\affiliation{ 
Université Lyon 1, CNRS, Institut Lumière Matière, UMR5306, F-69100, Villeurbanne, France
}%
\author{Franck Rabilloud}%
\affiliation{ 
Université Lyon 1, CNRS, Institut Lumière Matière, UMR5306, F-69100, Villeurbanne, France
}%
\author{Victor Despr\'e}%
\email{victor.despre@univ-lyon1.fr}
\affiliation{ 
Université Lyon 1, CNRS, Institut Lumière Matière, UMR5306, F-69100, Villeurbanne, France
}%

\title{
Supplementary material:~\\Real-time simulation of charge migration within the time-dependent Kohn-Sham DFT
}

\date{\today}

\onecolumngrid

\maketitle
\clearpage
\newpage

\begin{figure*}
\includegraphics[width=0.75\textwidth]{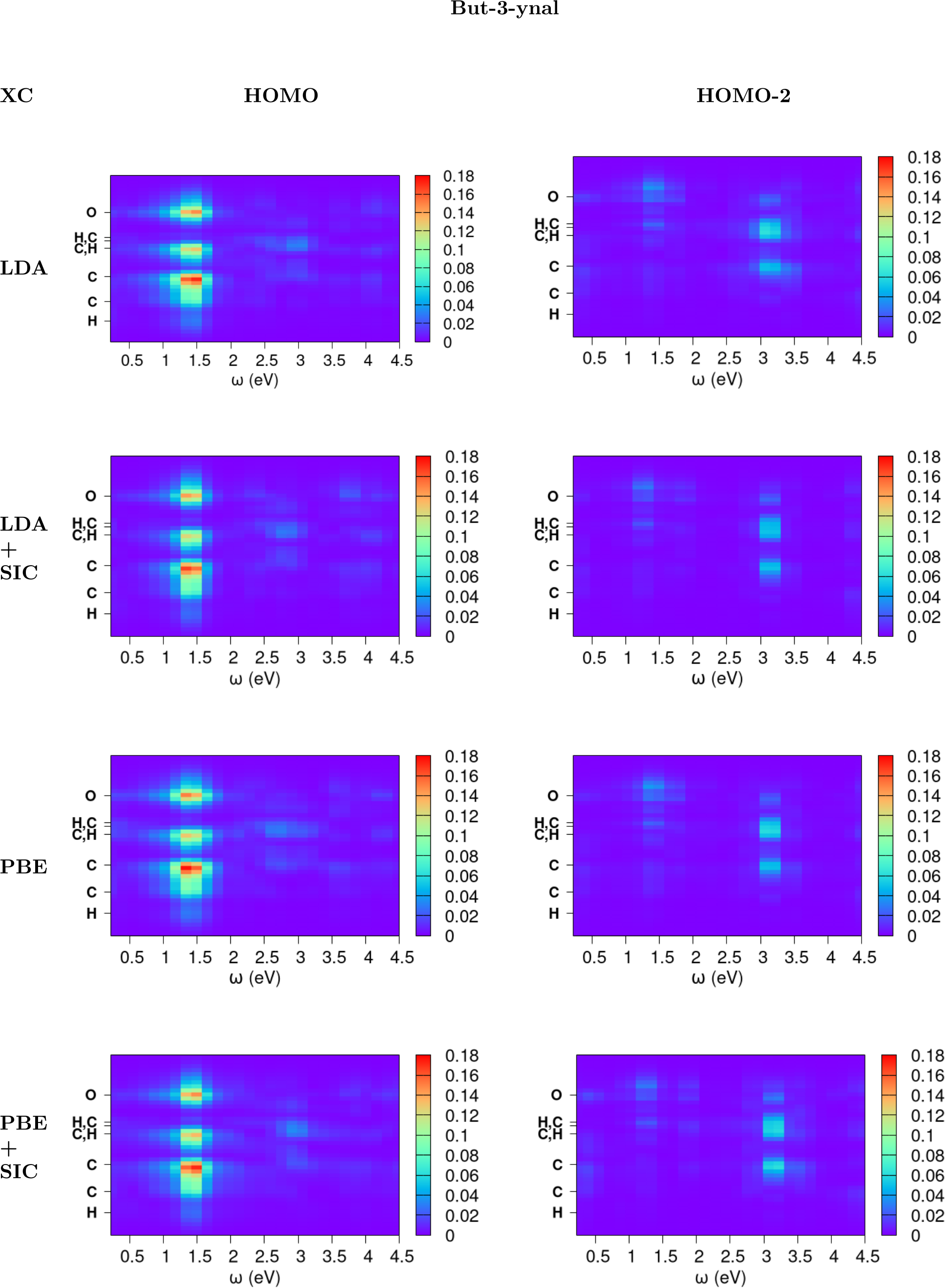}
\caption{
Fourier transform of the time-dependent density shows energy dependence of the dynamics of ionization by sudden by removal of an electron HOMO (left) and HOMO-2 (right) in but-3-ynal from different levels of approximation. 
}
\label{FT-but3ynal}
\end{figure*}

\begin{figure*}
\includegraphics[width=0.75\textwidth]{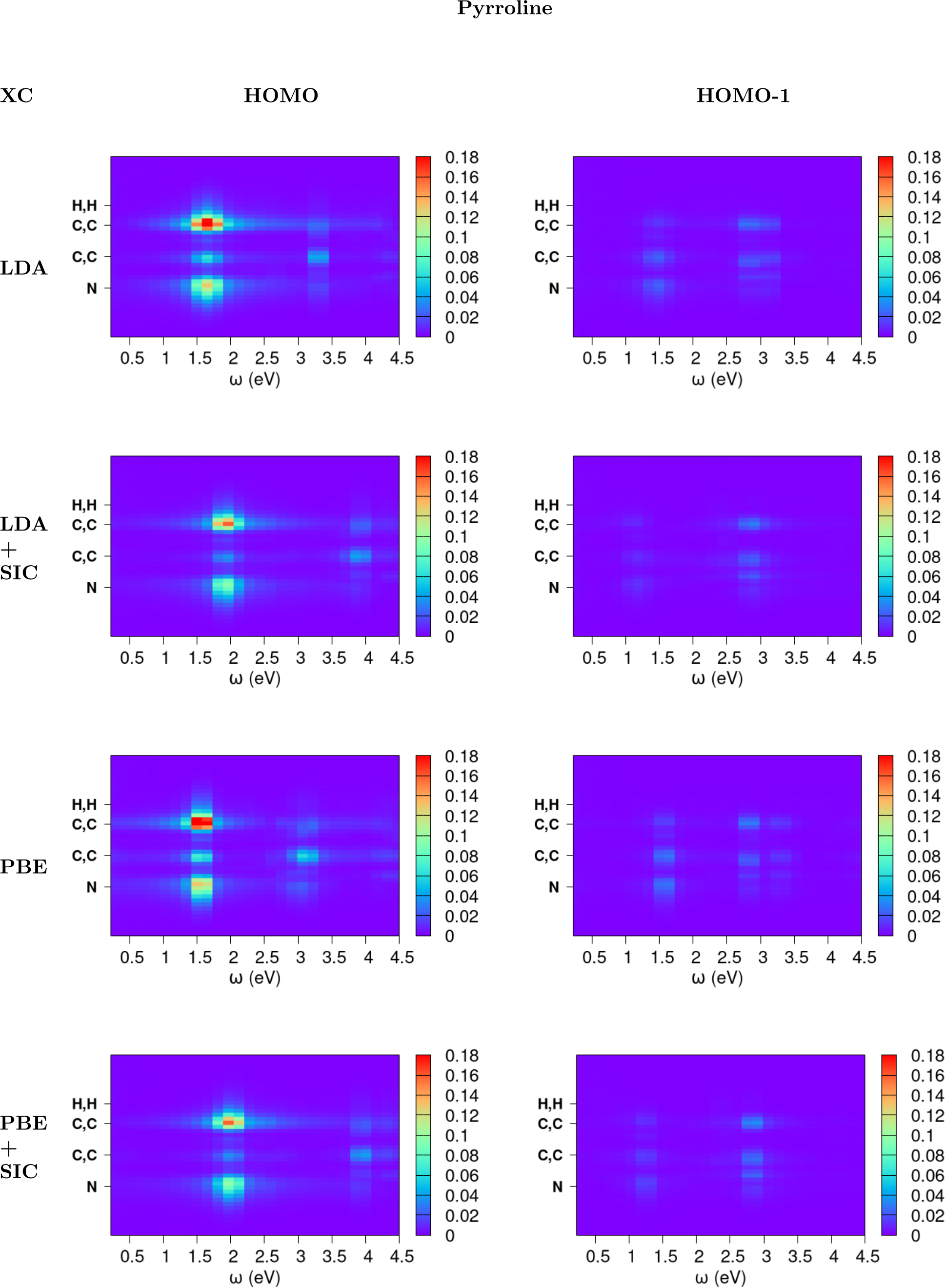}
\caption{
Fourier transform of the time-dependent density shows energy dependence of the dynamics of ionization by sudden by removal of an electron HOMO (left) and HOMO-2 (right) in pyrroline from different levels of approximation. 
}
\label{FT-pyrroline}
\end{figure*}

\begin{figure*}
\includegraphics[width=0.75\textwidth]{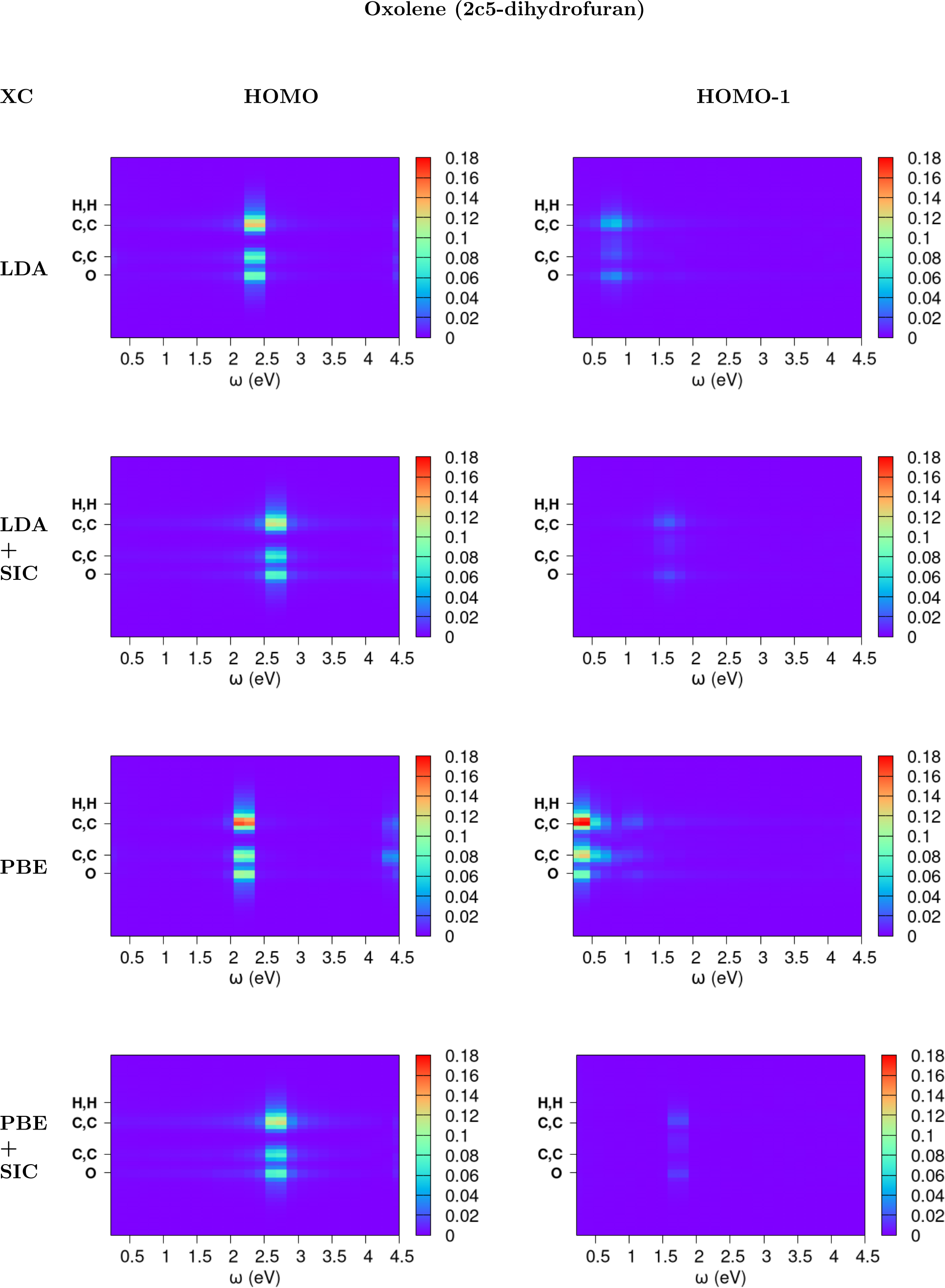}
\caption{
Fourier transform of the time-dependent density shows energy dependence of the dynamics of ionization by sudden by removal of an electron HOMO (left) and HOMO-2 (right) in oxolene from different levels of approximation. 
}
\label{FT-oxolene}
\end{figure*}

\begin{figure*}
\includegraphics[width=0.35\textwidth]{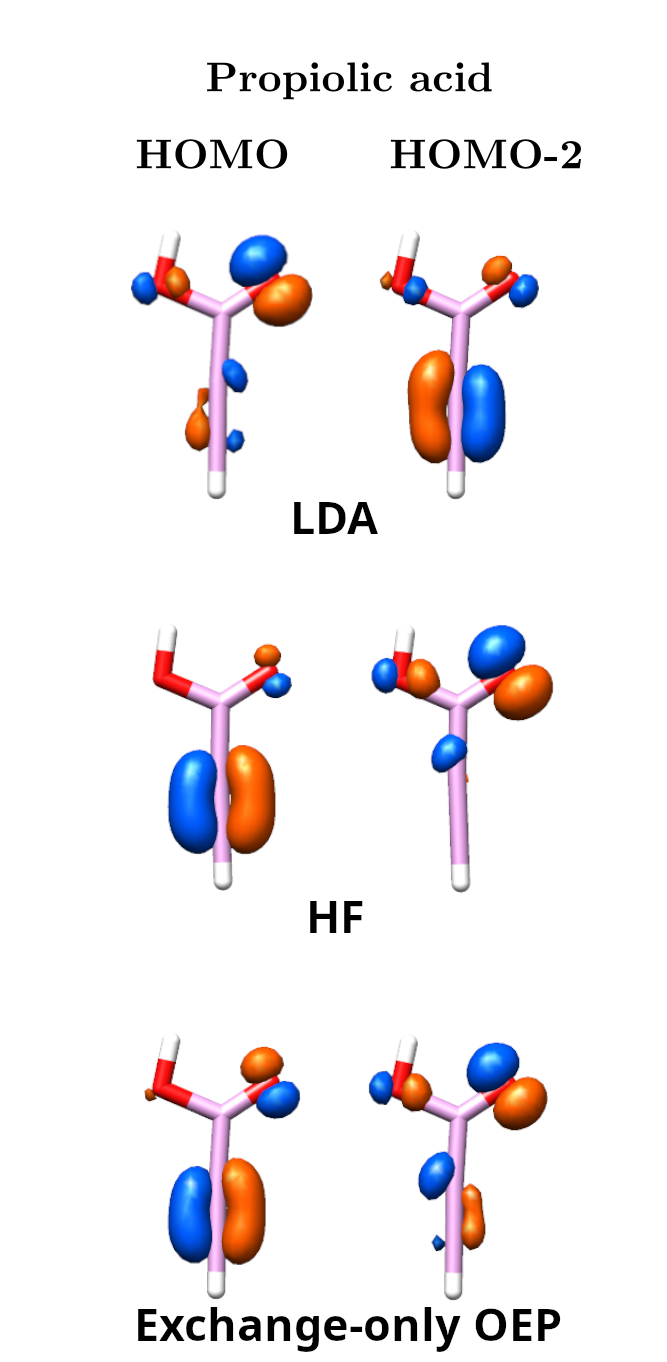}
\caption{
The orbitals HOMO and HOMO-2 which participate in the hole-mixing structure in propiolic acid are calculate in three different theory levels : local-density approximation for the XC functional, Hartree-Fock, and exchange-only optimized effective potential (OEP).
}
\label{orbitals_prop-acid_diff-theory}
\end{figure*}